# Identification of Opinion Leaders in a Telegram Network of Forwarded Messages

Giulia Tucci
Federal University of Rio de Janeiro (UFRJ) | Brazilian Institute of Information in Science and Technology (IBICT)
Rio de Janeiro, Brazil
giuliatucci@gmail.com

## ABSTRACT

Unraveling the role of opinion leaders in the digital realm, this study investigates the influence of key actors on Telegram, a hybrid platform that combines messaging app features with social network dynamics, where channel administrators gain a unique authoritative role. This research aims to create a method to identify opinion leaders in a network of forwarded messages on Telegram, adapting a method originally developed to be applied to Twitter. The adapted method is showcased through a case study during the 2022 Brazilian Presidential Election, involving the monitoring of 25 pro-Bolsonaro groups. The findings contribute to understanding the dynamics of digital opinion leadership, particularly in politically charged environments.

## 1 Introduction

Opinion leaders are individuals or entities that have significant influence over the attitudes and behaviors of others within a community or social group. They are often recognized for their expertise, credibility, and ability to shape public opinion in specific areas [10]. The role of opinion leaders has transcended traditional limits with the rise of digital platforms, becoming a multidimensional and complex phenomenon.In the networked society, with the establishment of social media platforms, this influence expanded, allowing politicians, journalists, celebrities, and other key actors to reach wider and more diverse audiences. Social media have become strategic spaces for opinion leaders who seek to extend their influence in the digital environment.

In the political context, opinion leaders play a crucial role in shaping and directing public discourse. A study analyzed Twitter data to investigate the behavior of opinion leaders who participated in political talk shows during the general elections of November 2019 in Spain [11]. The research revealed that messages about electoral results and media content predominate in the thematic agenda and that opinion leaders use Twitter to freely express their positions, especially negative ones, and to foster dialogue with users. The influence of opinion leaders on social networks can have a significant impact on electoral processes, as they can shape the public's perception of candidates and issues, thus influencing the outcome of elections [5]. Although social networks offer unprecedented opportunities for the dissemination of ideas, they also present challenges. Responsibility, ethics, and the veracity of the information shared are critical issues that opinion leaders must face in an increasingly polarized and fragmented online environment [20].

Messaging applications have consolidated as a means of daily communication [17] among people who relate with intimacy and/or trust [8]. This closeness and familiarity often lead users to trust the information shared in these environments, making significant the impact of these platforms on the formation of user opinion.

Telegram is a hybrid platform that combines the user interface architecture and functionalities of a messaging app with social network dynamics and environments – supergroups, for public discussion among thousands of members, and channels, for information broadcasting to unlimited audiences. Within the Telegram ecosystem, the communication dynamics established in a channel confer unique authority and legitimacy to its owner. By managing this channel, they assume the role of the entity responsible for transmitting information to a specific audience of subscribers. This prominent position grants the owner or administrator a recognized and respected voice, giving them an authoritative character in the context of the channel.

Developing a method to identify opinion leaders in Telegram is essential because this platform, with its 700 million users and status as the ninth most used worldwide [21], has an architecture and affordances that favor the dissemination of misinformation and hate speech [13,19]. Understanding the key influencers and their roles helps in comprehending how information and opinions form and spread in digital communities, which is critical for devising strategies to counter misinformation and promote positive discourse in these influential digital spaces.

The aim of this work is to develop a consistent process for identifying which entities act as opinion leaders in a network of forwarded messages on Telegram, based on the method developed by Rehman, Jiang, et al. [18] with data from Twitter.

## 2 From Twitter to Telegram: adapting the method

In this work, I adapted a method developed to be applied to Twitter datasets to identify the opinion leaders on a specific event or topic. The researchers justify the use of Twitter because the platform has a "fast information flow with a huge impact on the formation of public opinion" [18]. My objective is to adapt this method for a dataset extracted from Telegram – a network of forwarded messages obtained from the monitoring of public groups. This network was built from the intense circulation of forwarded messages among thousands of users.

Telegram offers two different public interactions environments where the information flow dynamics differ significantly: broadcasting channels and discussion groups. In groups, all members have the privilege to send messages and participate in discussions, generally resulting in a broader and more horizontal circulation of information. In groups, there is no centralized figure with exclusive authority; they potentially constitute a collaborative environment where participation and diversity of perspectives shape the flow of information. On Twitter, a retweet functions as a method of reporting speech and disseminating the discourse of other users, that is, it is a form of endorsement of the original content [6].

Nobari, Sarraf and colleagues [15] illustrate the relationship between a retweet and the forwarding of messages between channels as methods of confirming a message. Consequently, based on this understanding, I propose a definition for key users in the Telegram message forwarding network, drawing upon the definition formulated by Rehman, Jiang, et al., [18] for the retweet network. Since messages forwarded to individual users are private and unextractable, they are not deemed potential opinion leaders. However, it is plausible that individual users might act as opinion leaders in the network, although this scenario is not considered here. Furthermore, this investigation aims to identify key users leading debates within a Telegram sphere, focusing exclusively on broadcasting channels as potential opinion leaders.

### 2.1 Identification of key Users in a Telegram network

On Twitter, when a user retweets content tweeted by another (or by themselves), the resulting post is made in their profile feed and, if their profile is public, it may appear in the public feed. This retweet creates a connection between the user who retweeted and the one who was retweeted. Thus, a directed network relationship is formed between the users, and in this network, the direction of the bond representing this relationship points from the author of the retweet (who performed the action of retweeting) to the author of the tweet that was retweeted.

On Telegram, when an administrator forwards a message to a channel, the message will appear in that channel's feed and display the original author's information (source of the forwarded message), as well as the name of the channel itself, the author of the message forwarding action. If a subscriber in this channel then forwards the message to their personal contacts, the message will carry the original authorship information but will omit the secondary forwarding source information (the channel through which the user accessed the content).

The action of forwarding a message on Telegram creates not only a relationship between the person who executed the forwarding function and the author of the forwarded message but also connects the environment to where the message was forwarded and its author. This last relationship, the recipient-author of the forwarded message relationship, is the one investigated here. The resulting network is also a directed network, and to represent the real direction of the flow of information, the bond representing the message forwarding points from the message's source (author of the forwarded message) to the channel or group to where it was forwarded. The message forwarding author is not represented. Although the person who performs the forwarding action is not represented in the network, in the case of channels, it is known that when they receive forwarded messages (when forwarded messages are posted in their feed), the action was performed by the administrators themselves.

On Twitter, the conversation starter node in the retweet network has a high in-degree and low out-degree [18], that is, when users want to talk about a topic, they retweet content from this user, and the same does not retweet content from others on this topic. On Telegram, the conversation starter is the one whose messages are widely forwarded and who forwards few (if any) messages from other entities. Thus, the node of the conversation starter in the Telegram message forwarding network has a high out-degree and low in-degree. Analogous reasoning is used for the active engager, who has a large presence of content forwarded from other sources and low (or none) of their own content forwarded to other channels. Below, I propose a definition of the five types of key users for message forwarding to Telegram channels:

1. **Conversation starter**: Has their messages forwarded to other channels but receives a relatively low number (or none) of forwarded messages (administrators do not usually forward messages from other Telegram sources to this channel).
2. **Influencer**: receives a relatively high number of forwarded messages (administrators forward messages from other Telegram sources to this channel) and has their messages frequently forwarded to other channels.
3. **Active engager**: Frequently receives forwarded messages from other channels (administrators forward messages from other Telegram sources on this channel) but is not usually a forwarding source (rarely has their content forwarded to other channels).
4. **Network creator**: a Telegram channel that connects two or more influencer channels.
5. **Information bridge**: a Telegram channel that connects an active engager channel to an influencer channel.

Figure 1 presents a diagram showing the association links between key-users figuring in a Telegram forwarded messages network.

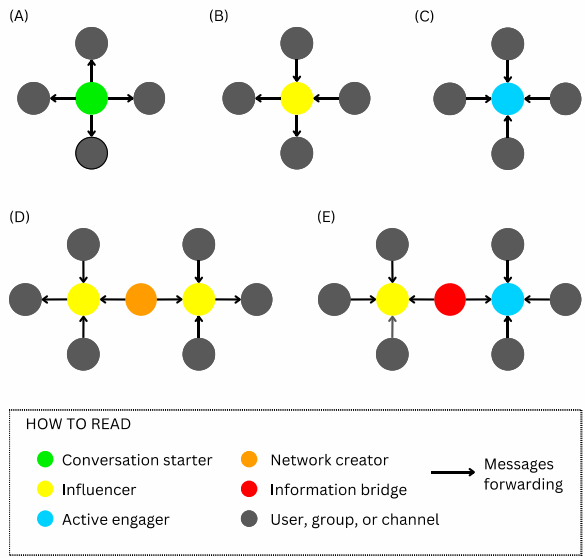

**Figure 1: Schematic representation of forwarding messages links between key users in a Telegram forwarded messages network.**

## 3 Case study: identification of opinion leaders in a network of electoral campaigners in Telegram

In this section I present an experimental analysis conducted during August 2022, explaining the dataset and the methodology employed.

### 3.1 Introduction

The 2022 presidential elections in Brazil unfolded amidst significant political polarization and contested narratives [4]. The Brazilian Electoral Court focused on implement measures to ensure the stability and truthfulness of online discourse [7]. Facing resistance from Telegram, the court implemented a blockade, compelling the platform to adopt measures against misinformation [14]. However, these measures targeted only broadcast channels, with no actions directed at public groups.

Considering the importance of monitoring content in public Telegram groups, I selected 25 groups supporting the then-incumbent president and re-election candidate.

### 3.2 Data and Methods

*3.2.1 Dataset description.* The dataset was created in two steps. First, I extracted all messages shared on 25 monitored groups during August 2022, month of the beginning of the electoral campaign in Brazil. In this first step, I extracted 195.567 messages using the 4CAT tool to access the Telegram API [16]. Then, based on the messages metadata, I filtered the forwarded messages and listed the usernames of the forwarded content sources. To ensure the thematic relevance of the entities in the Telegram network, I focused the analysis on sources that appeared at least 50, resulting in 241 users (94), groups (3) and channels (142).

In the second step of data extraction, I used 4CAT [16] to extract all messages posted in August 2022 by these 145 Telegram channels and groups, resulting in a total of 91,682 messages, which, combined with the first step collection, amounted to 287,249. The data were again filtered to include only messages forwarded by entities with public usernames and to identify the sources, totaling 80,508 forwarded messages.

*3.2.2 Network Analysis.* The forwarded messages network represents the connections between the groups and channels to where messages were forwarded and the source of these messages (channels, groups, or users). In other words, it represents the network relationship between groups and channels that received forwarded messages posted by other channels, groups, and users. It should be noted that the network does not carry information about who performed the forwarding action (which would be the author of the forwarded message), but rather the author of the original message.

The data was processed with Gephi [1] to create and analyze the network. The pro-Bolsonaro forwarded messages network is a direct network consisting of 2.517 nodes and 9,198 edges. The nodes represent channels, groups, or users thar created or received messages that were forwarded and the edges, the links that connect the source and the receptor of each of the 80,508 forwarded messages.

### 3.3 Results

The key users who emerged as opinion leaders in the pro-Bolsonaro Telegram forwarded messages network are listed in Table 1, which is ordered, in descending frequency, by the frequency of the channels in the network.

Figure 2 displays two versions of the forwarded messages network, with the nodes distributed spatially by two different algorithms, the OpenOrd layout [12] (Figure 2-A) and the Yifan-Hu layout [9] (Figure 2-B). The Figure indicates the positioning of the identified key users in the forwarded messages network. In both network configurations, the size of the nodes was varied based on betweenness centrality and the color based on the type of user in the network, according to the information in the legend. The labels of the nodes, except for those of the identified key users, were hidden to facilitate visualization.

In Figure 2-A, the network was built with the OpenOrd algorithm [12], which forces the formation of communities and distances these communities from each other, and in Figure 2-B, with the Yifan-Hu algorithm [9], which positions nodes with weaker connections to more outer regions of the network, distancing them from more strongly connected nodes.

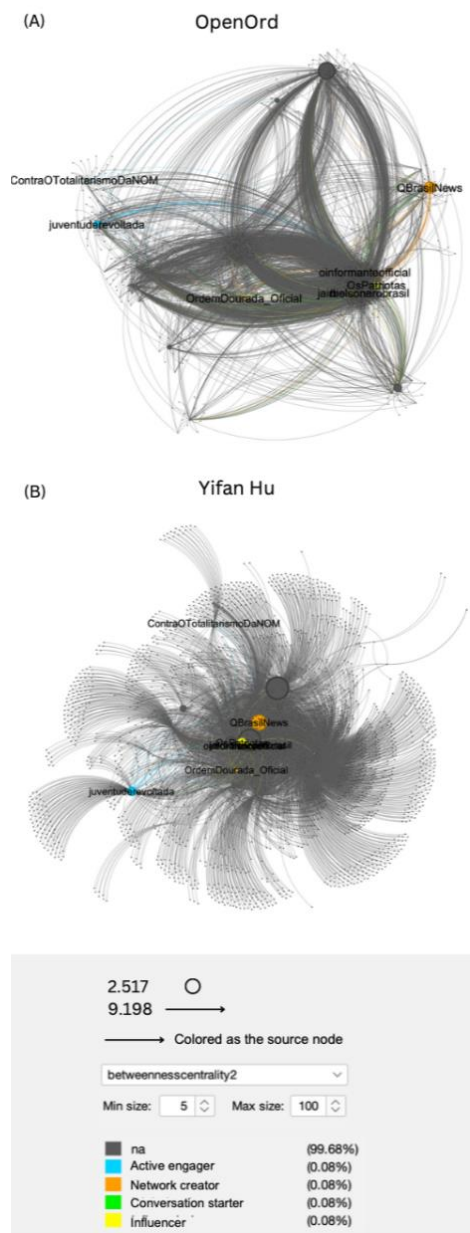

**Figure 2:** Key users' location on the forwarded messages network. (A) Open-Ord layout. (B) Yifan-Hu layout.

Figure 3 demonstrates the pattern of interaction among the entities that comprise the pro-Bolsonaro message forwarding network on Telegram in two spatial configurations of the nodes, one created with the OpenOrd algorithm (Figure 3-A) and the other with the Yifan Hu (Figure 3-B). The calculation of the network's modularity in Gephi [2] produced a coefficient of 0.383 and identified a total of ten communities, meaning there are ten distinct groupings among users, groups, and/or channels that have more similar communication patterns among themselves when it comes to message forwarding. To facilitate the identification of the formed groupings, in both network visualizations, the nodes were colored according to the community to which they belong, and their size was fixed based on their degree.

**Table 1: Key users of the pro-Bolsonaro forwarded messages network on Telegram during August 2022**

| Channel | Type | f | i.d. | o.d | BC |
|---|---|---|---|---|---|
| jairbolsonarobrasil | Conversation starter | 1491 | 0 | 38 | 0 |
| OsPatriotas | Influencer | 1373 | 12 | 29 | 136,142,893 |
| QBrasilNews | Network creator | 919 | 47 | 24 | 22562.4203 |
| juventuderevoltada | Active engager | 916 | 130 | 16 | 122,787,921 |
| OrdemDourada_Oficial | Network creator | 757 | 57 | 21 | 35,514,004 |
| oinformanteofficial | Conversation starter | 551 | 4 | 38 | 1,567,333 |
| ContraOTotalitarismoDaNOM | Active engager | 461 | 41 | 4 | 16,253,078 |
| bielconn | Influenciador | 349 | 26 | 30 | 92,008,166 |

Source: calculations developed in Gephi [1]. Title line legend: Channel – channel's username (@), Type – type of key user, f – frequency, i.d. – in-degree, o.d. – out-degree, and BC – betweenness centrality.

## 3.4 Discussion

The conversation starters are channels that have their messages used by other channels as a source of information, but use little (if any) forwarded content from other channels to disseminate content. In other words, other channels in the network tend to post content in their feed that is forwarded from the conversation starter channels. Channels acting as conversation starters in the pro-Bolsonaro Telegram network include Bolsonaro's own channel and a conspiracist channel (@oinformanteoficial). Both have an out-degree of 38, the highest in the network, meaning they acted as the source of forwarded messages for 38 entities within the network. During August 2022, these were the most common channels to start a discussion on a specific topic, and the content posted on their channels served as a trigger to initiate new information flows in the network. Other centrality measures are displayed in Table 1. It's expected that during his reelection campaign, and with a network built from a sample of groups supporting his campaign, Jair Bolsonaro would be the main conversation starter in the network.

Identification of Opinion Leaders in a Telegram Network of Forwarded Messages

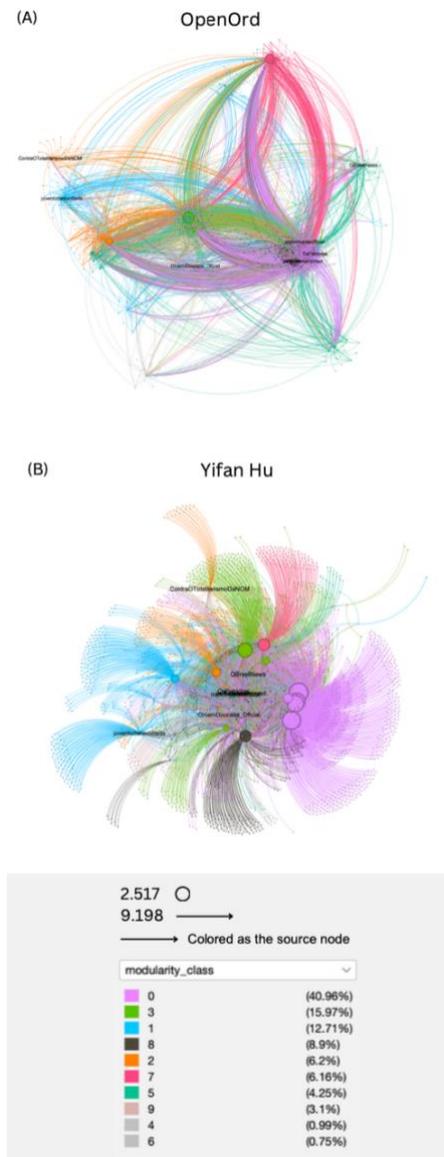

**Figure 3: Communities detected on the forwarded messages network. (A) Open-Ord layout. (B) Yifan-Hu layout.**

Channels in the active engager category are those that commonly post forwarded messages in their feed, but their content does not resonate and is often not forwarded to other entities. The nodes representing these channels have a high in-degree compared to their out-degree. The junk news channel @juventuderevoltada, with an in-degree of 130 (and out-degree of 16), is the main conversation starter in the network, followed by the conspiracy channel @ContraOTotalitarismoDaNOM (in-degree 41 and out-degree 4). Although not having the highest in-degree, active engagers are those with the highest in-degree/out-degree ratio in the network. They are classified as key users because they play an important role in forwarding messages in the network, but as their messages are seldom forwarded, they cannot be considered opinion leaders (REHMAN et al., 2020).

Influencers are channels with abundant in- and out-degree connections, meaning they forward messages from other entities to their feeds and are used by other network entities as a preferred content source. The channels @bielconn and @OsPatriotas, with the highest out-degree and a close number of in-degrees (Table 1), act as opinion leaders. They generate an impact on the network, as their content is forwarded by other entities and they publish messages forwarded from other sources in their feeds. Additionally, influencer channels connect isolated entities, which have no connection with the rest of the network [18].

A channel that acts as a network creator is one that, in addition to having numerous out-degree connections and some in-degree, connects two influencers. This important role is played in the network by the channels @QBrasilNews and @OrdemDourada_Oficial, which connect the channels @bielconn and @OsPatriotas. No channels acting as information bridges were identified in the message forwarding network, similar to how Rehman, Jiang, and colleagues [18] did not identify this type of key user in the retweet network they analyzed.

In both network configurations, the active engagers, @ContraOTotalitarismoDaNOM, and @juventuderevoltada, which act by disseminating content in the network through forwarding messages from other sources to their channel audiences, are positioned in more peripheral clusters. These two channels, despite not being opinion leaders in the network, have some relevance in the debate, as their content is forwarded by other network entities (see out-degree greater than zero in Table 1).

In Figure 2-A, another key user appears further from the center of the network, the conspiracy network creator channel @QBrasilNews, unlike Figure 2-B where all, except the active engagers, appear in the center of the network. The influencers and conversation starters are positioned in the most central and connected region in both OpenOrd and Yifan-Hu, as they are the most relevant sources of information in the network and the influencers tend to create a larger volume of content in order to influence the opinion of users in the network. The active engagers and network creators, on the other hand, are channels that act as a bridge for the information circulating in the network through forwarding content.

There is a presence of entities with low degrees connected to opinion leaders in the network, which ensures that these entities have access to up-to-date content posted by the most influential channels. It is important to highlight that many users were a source of messages for the monitored groups and, despite users also being potential recipients of forwarded messages, they will be connected exclusively to the nodes of channels and groups to where they forwarded a message (thus being a source of forwarding). This occurs because regular users do not have a public environment from which circulated messages can be collected (in addition to the need to preserve data for ethical and legal reasons – user privacy).

Considering the clustering tendency of the network, the low modularity coefficient suggests that, although detectable communities exist, the network is not strongly segmented into isolated groups. There is a certain mix and flow of information between different communities, which may indicate a more open and interconnected network. Although the existence of distinct communities might represent different currents or interests within the pro-Bolsonaro network, the low coefficient indicates that the forwarded messages are not being confined to specific sets of Telegram entities. Two active engagers make up distinct communities and appear further from the center of the network, @juventuderevoltada (community 1, in blue) and @ContraOTotalitarismoDaNOM (community 2, in orange).

## 3.5 Limitations and ethical considerations

This work has the following limitations. First, the number of groups chosen as seed (25 groups). Despite starting the investigation with a relatively small sample of groups, it was possible to construct a robust dataset. Unlike Telegram channels, the possibility of message exchange among thousands of group members tends to facilitate the generation of a high volume of content. Second, a notable aspect of Telegram is that not all shared content in the seed groups may have been captured. This uncertainty arises due to Telegram's self-destruct message feature, where messages are automatically and permanently deleted after a predetermined period. Additionally, this study did not account for potential secret discussion groups or private channels disseminating pro-Bolsonaro content.

Ethically, the study adhered to strict guidelines: data were sourced exclusively from public channels and groups, with group member identities anonymized. Usernames mentioned in this chapter are associated with public channels and groups, constituting public information accessible via Telegram's search functionality.

## 4 Conclusion

This investigation revealed aspects of the structure that supports the dynamics of information circulation on Telegram. Through the application of algorithms and network analysis techniques, different types of opinion leader channels were identified: influencers, conversation starters, active engagers, and network creators. Understanding the roles of these actors in the network provides valuable insights into how information is disseminated. The analysis revealed a network structure with detectable communities, but also with a certain permeability and flow of information among these communities.

Using the adaptation of the definition of the five types of key users, originally identified in a network built with Twitter data [18], this chapter formulated a guide to identify channels that emerge as opinion leaders in networks built with Telegram data. This approach can be adopted by other researchers seeking to explore and understand the nuances of interactions and influences on Telegram, thus contributing to the field of study of digital social platforms.

The distinction between channels and groups on Telegram brings to light significant issues about the power of influence, the dissemination of information, and the formation of opinions within these different communicational contexts. This distinction is not just a technical feature but reflects a complex interaction of social, cultural, and technological factors. Based on Danah Boyd's concept of networked publics [3], it is observed that Telegram's structure reorganizes the flow of information and the way people interact with information and each other. Networked publics are restructured by network technologies, representing simultaneously a space and a set of people. This restructuring amplifies and complicates traditional notions of publics, emphasizing the interconnection and accessibility of spaces and information.

On Telegram, this network structure facilitates the dissemination of misinformation, as well as the formation and acting of social identities. Its hybrid character, where channels act as public environments to transmit a user's messages and groups serve as community platforms, creates a unique environment that allows both the consumption and production of cultural objects, as well as the creation and dissemination of media.

A notable limitation of this study lies in the need to consider ethics and privacy in data collection, especially in relation to regular users who do not have a public environment from which messages can be collected. This restriction, although essential for preserving individual privacy, hides a part of the network and potentially limits the complete understanding of the message forwarding dynamics within the pro-Bolsonaro community on Telegram.

The distinction between channels and groups on Telegram brings up relevant questions about the power of influence, the dissemination of information, and the formation of opinions within these different communicational contexts. For example, it would not be possible to develop an equivalent analysis to investigate opinion leaders in a network of forwarded messages in public WhatsApp groups. Firstly, because WhatsApp does not display the information of the original message's author, only of the user who executed the forwarding. Furthermore, on WhatsApp, it would be necessary to obtain sensitive user information like phone numbers, as the tool does not offer the possibility of creating a username. Thus, once again, there is evidence of the multiple character of Telegram and the influence of this character on the flow of information and how this hybridity facilitates the dissemination of misinformation.

ACKNOWLEDGMENTS

This study was financed by the Coordenação de Aperfeiçoamento de Pessoal de Nível Superior - Brasil (CAPES) - Finance Code 001